\documentclass[%
reprint,
％preprint,
 amsmath,amssymb,
 aps,
]{revtex4-2}

\usepackage{graphicx}% Include figure files
\usepackage{dcolumn}% Align table columns on decimal point
\usepackage{bm}% bold math
\usepackage{booktabs}
\usepackage{color}

\begin{document}

\title{Evolutionary Innovation by Polyploidy}% Force line breaks with \\

\author{Tetsuhiro S. Hatakeyama}
\email{hatakeyama@elsi.jp}
\affiliation{%
Earth-Life Science Institute (ELSI), Tokyo Institute of Technology, 2-12-1-IE-1 Ookayama, Meguro-ku, Tokyo 152-8550, Japan
}%

\author{Ryudo Ohbayashi}
\affiliation{
Department of Biological Sciences, Tokyo Metropolitan University, 1-1 Minami-Osawa, Hachioji-shi, Tokyo 192-0397, Japan
}%

\date{\today}% It is always \today, today,

\begin{abstract}
The preferred conditions for evolutionary innovation is a fundamental question, but little is known, in part because the question involves rare events.
We focused on the potential role of polyploidy in the evolution of novel traits.
There are two hypotheses regarding the effects of polyploidy on evolution:
Polyploidy reduces the effect of a single mutation and slows evolution.
In contrast, the gene redundancy introduced by polyploidy will promote neofunctionalization and accelerate evolution.
Does polyploidy speed up or slow down evolution?
In this study, we proposed a simple model of polyploid cells and showed that the evolutionary rate of polyploids is similar to or much slower than that of haploids under neutral selection or during gradual evolution.
However, on a fitness landscape where cells should jump over a lethal valley to increase their fitness, the probability of evolution in polyploidy could be drastically increased, and the optimal number of chromosomes was identified.
We theoretically discussed the existence of this optimal chromosome number from the large deviation theory.
Furthermore, we proposed that the optimization for achieving evolutionary innovation could determine the range of chromosome number in polyploid bacteria.
\end{abstract}

\maketitle

%\tableofcontents

\section{Introduction}

How novel traits in organisms have emerged is one of the most critical questions in biology.
Through decades of theoretical development in population genetics \cite{Fisher1930, Crow1970, Hartl1997} and recent large-scale data analysis and laboratory evolutionary experiments \cite{Conrad2011, Palmer2013, Maeda2020}, we have learned a great deal about the quantitative evolution of traits.
However, the mechanism of evolutionary innovation is challenging to investigate because we must consider the probability of rare events occurring.
In this aim, we consider the conditions under which novel traits are likely to evolve from existing organisms believed to have been more likely to acquire evolutionary innovation.

We first focus on polyploid bacteria, particularly cyanobacteria.
Cyanobacteria are more diverse regarding their morphology and habitats than most other bacteria \cite{Flores2008}.
They range from unicellular cocci and rods to bead-like and branching multicellular forms that exhibit cell differentiation, and their habitats range from underwater to in the soil and above the ground.
This indicates that cyanobacteria are more likely to achieve evolutionary innovation than most other bacteria.
One of the significant differences between cyanobacteria and other bacteria, other than photosynthesis, is their polyploid nature \cite{Griese2011}.
Interestingly, most marine cyanobacteria, although photosynthetic, are known to be haploid and possess a simple unicellular morphology \cite{Griese2011}.
Such polyploidy has also been reported in halophilic archaea living in extreme environments \cite{Zerulla2014}.
Although why these species have been able to diversify and migrate to extreme environments is unclear, the diversity of polyploid bacteria and archaea implies the potential impact of polyploidy on the evolution of novel traits.
Although several studies examined the physiological significance of polyploidy in bacteria \cite{Ohbayashi2019, Paijmans2016}, the evolutionary impact of polyploidy has rarely been studied.

The impact of polyploidy on evolutionary innovation has also been demonstrated in medicine in recent years.
For example, cell wall-deficient/defective bacteria, i.e., the L-form of bacteria, which are known to be polyploid \cite{Briers2012}, were reported to easily develop resistance to multiple drugs \cite{Domingue1997, Leaver2009}.
Furthermore, in the last decade, it has been intensively reported that tumor cells become large and polyploid through endoreplication or cell fusion under stress conditions \cite{Zhang2014, Song2021}.
Such cells are termed polyploid giant cancer cells (PGCCs). 
These cells are recognized to be important for the development of resistance to various cancer therapies, e.g., chemotherapy, hormone therapy, immunotherapy, and systemic radiation therapy, and for metastasis, i.e., for achieving novel traits.
Although the need to understand the influence of polyploidy on evolution is increasing even in the medical field, few results have been obtained.

How can polyploidy actually promote the evolutionary innovation? 
There are two conflicting hypotheses.
In polyploidy, even if favorable or unfavorable mutations are introduced into one genome, the effect will be weakened by the presence of a large number of unmutated genomes \cite{Koch1984}, and the resulting phenotypic changes will be smaller than those in haploidy.
According to Fisher's fundamental theorem of natural selection \cite{Fisher1930}, the evolutionary rate should be proportional to the genetic variance of the fitness, and the evolutionary rate should be slower in polyploid organisms because the genetic variance of the phenotype is small in such organisms.
A similar point was raised theoretically in a previous study as the minority control of genetic information \cite{Kaneko2002}.
By contrast, polyploidy is the same situation as when whole-genome duplication occurs.
Such whole-genome duplication was postulated by Susumu Ohno as a source of neofunctionalization attributable to the redundancy of duplicated genes \cite{Ohno1970}.
According to this idea, the evolution of novel traits appears more likely to occur in polyploidy.
Is one of these hypotheses correct, or can both apparently contradictory hypotheses coexist?

In this paper, we first proposed one of the simplest models of the evolution of polyploid organisms. 
We demonstrated that the evolutionary rate in polyploidy is at best unchanged or is much slower than that in haploidy under neutral selection or on the smooth fitness landscape.
This is because, as mentioned previously, the genetic variance of the phenotype is lower in polyploids than in haploids.
However, the evolutionary dynamics of the same model were very different in the multimodal fitness landscape, where cells should jump over a lethal valley to increase their fitness.
Although polyploidy still reduced the genetic variance of the phenotype, the probability of evolving a novel trait increased with increasing ploidy levels, and an optimal number of chromosomes existed.
We demonstrated with the aid of the large deviation theory that the evolution of such novel traits could be attributed to a bias in genetic information on chromosomes in polyploidy.
Furthermore, we hypothesized that the range of genome number in various cyanobacterial species results from the optimization of chromosome bias and hence the evolutionary innovation.

\section{Model}

We considered a simple case in which phenotype depends on a translation product from a single locus on chromosomes.
That product is assumed to be transcribed and translated from all chromosomes at an equal rate, as experimentally observed \cite{Ohbayashi2019, Zheng2017}, and it has activity represented by a single scalar quantity, namely $x_{i,j}(t)$ for a $j$th chromosome in an $i$th cell at $t$th generation.
Alternatively, we can consider the case in which various genes on the chromosome contribute to a single trait, and the sum of their contributions is represented as a single scalar quantity.
Even in that case, it does not affect the conclusion.
Every cell has $N$ homologous chromosomes that carry somewhat different copies of the same genes; and thus, the genotype is given as a set of $x_{i,j}(t)$, and the phenotype $y_i(t)$ is given as a function of that set.
The fitness of $i$th cell is represented as $f(y_i(t))$ as a function of the phenotype.
Inherited chromosomes are mutated every generation by adding a random value as $x_{i,j}(t+1) = x_{i^\prime, j^\prime}(t) + \mathcal{N}(0, \sigma^2)$, where $\mathcal{N}(0, \sigma^2)$ is a random number sampled from the Gaussian distribution with a mean of 0 and variance of $\sigma^2$.
We set $\sigma$ as one unless otherwise noted.
For the sake of simplicity, we assumed that the phenotype is determined by the averaged activity of the product, i.e., the fitness is given by $y_i = \langle x_{i,j} \rangle_{\rm c}$, where $\langle z \rangle_{\rm c}$ is the average of $z$ over chromosomes given by $\langle z \rangle_{\rm c} = \Sigma_j z_j/N$.
We also defined $\langle z \rangle_{\rm p}$ as the average of $z$ over the population given by $\langle z \rangle_{\rm p} = \Sigma_i z_i/M$, where $M$ is the total population and set as 1000 in simulations.
In addition, $\langle z \rangle_{\rm ens}$ indicated the ensemble average among different samples.
We set the number of samples as 10,000 with a different random seed.

We investigated the extent to inheritance modes affect the evolutionary rate.
We defined two different modes of inheritance for polyploidy to investigate how such differences affect evolution.
We introduced two extreme modes: set inheritance and random inheritance modes (Fig. \ref{fig:neutral_selection}A).
The set inheritance mode is similar to the chromosome segregation in eukaryotes in that daughter cells inherit a set of chromosomes from a mother cell.
All of the chromosomes in the mother cell are precisely duplicated individually and inherited by daughter cells in the same manner.
Thus, duplication of the same chromosome does not occur.
Conversely, in the random inheritance mode, daughter cells randomly inherit the mother cell's chromosomes.
Each chromosome is randomly chosen and inherited until the number of chromosomes reaches the same number as that in the ancestor.
Thus, multiple copies of some chromosomes may be inherited, whereas other chromosomes may not be inherited at all.
Organisms lacking have sophisticated DNA replication control mechanisms, such as bacteria, are believed to have inherit chromosomes via the random inheritance mode.
At least, it was experimentally observed that polyploid bacteria replicate chromosomes in a one-by-one manner \cite{Ohbayashi2019} opposed to simultaneously as observed in eukaryotic cells.

Each cell is selected according to fitness, whereas each chromosome is neutrally selected in each cell because it usually has no fitness.
Thus, chromosomes in the population are only selected through selection at the cell level.

\section{Results}

\subsection{Evolutionary rate drastically depends on the mode of inheritance in polyploidy}

First, we considered that each cell is neutrally selected, and therefore, each chromosome is also neutrally selected.
In this situation, evolutionary speed can be simply defined as the rate of phenotype changes, i.e., the diffusion constant in the phenotypic space ($D$).
Because the phenotype exhibits Brownian motion in the phenotypic space, the diffusion constant is proportional to the variance of $y_i(t)$.
Thus, we must consider the dependence of the variance of phenotype on the mode of inheritance.

\begin{figure}[htbp]
\centering
\includegraphics[width=0.8\linewidth]{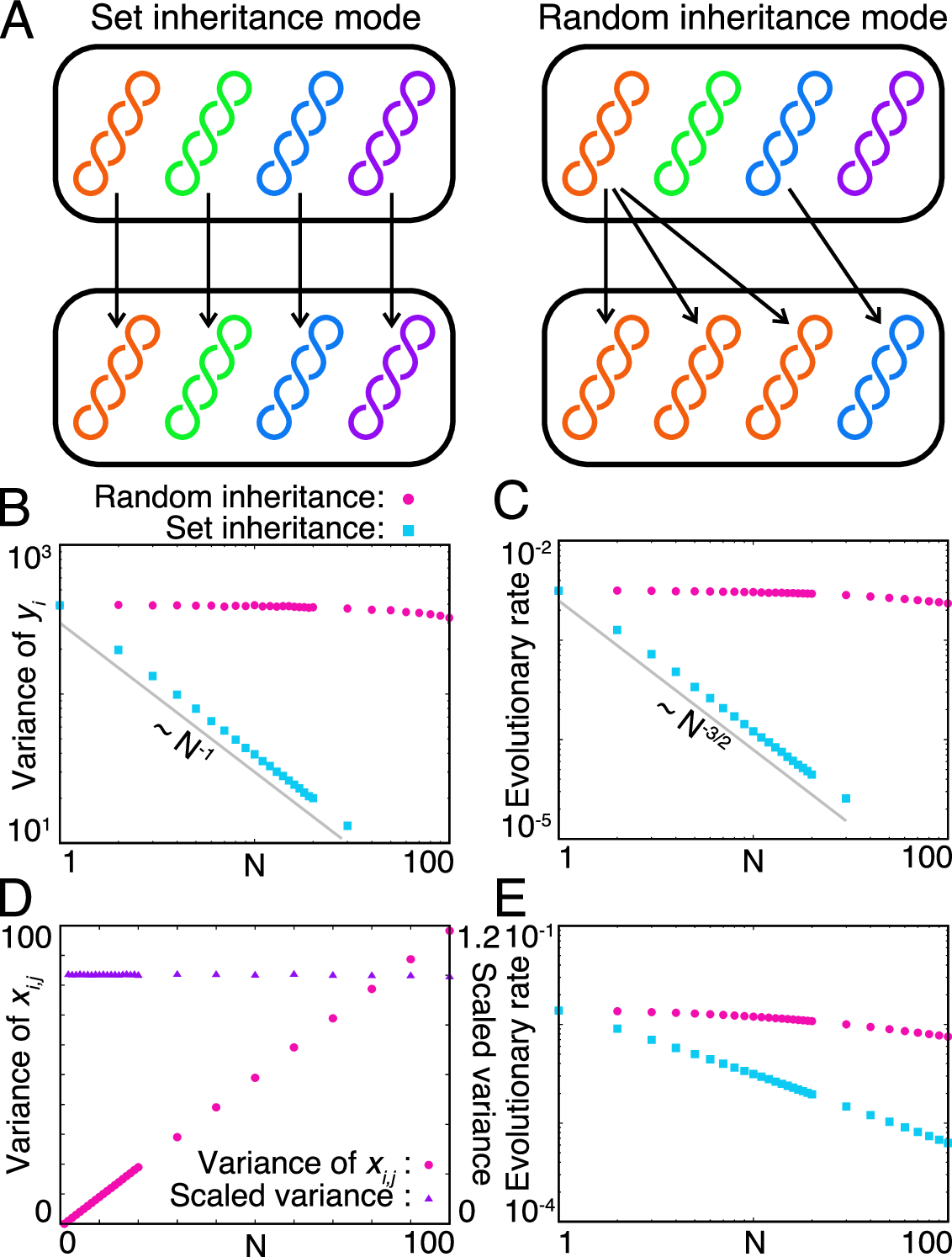}
\caption{
Dependence of evolution on modes of chromosome inheritance.
A) Scheme of modes of chromosome inheritance.
B and C) Dependence of (B) the variance of phenotype and (C) the evolutionary rate on the number of chromosomes.
The variance of the phenotype was calculated at time $t = 500$.
The evolutionary rate was calculated as the inverse time when an absolute value of the phenotype of a cell exceeded the threshold $y_{\mathrm{th}} = 50$.
Each point represents an average of 10,000 samples.
Magenta circles and cyan squares represent data for the random inheritance and set inheritance modes, respectively.
D) Dependence of the population average of variance among chromosomes with the random inheritance mode on the number of chromosomes.
Purple triangles represent the variance among chromosomes scaled by dividing by $N-1$.
E) Evolutionary rate on the smooth landscape ($f(y) = -2\cos(x/15)$).
We set $x_{i,j}(0) = 0$ as the initial condition and plotted $1/\langle \tau_t \rangle_{\mathrm{ens}}$, where $\tau_t$ is the time when the population average of fitness crossed the threshold value ($f^{\mathrm{th}}(y) = 1.5$).
\label{fig:neutral_selection}}
\end{figure}

In the set inheritance mode, each chromosome is independently mutated in the same manner, and thus, $x_{i,j}(t)$ has no correlation for different $j$.
In other words, each $x_{i, j}(t)$ is an independent and identically distributed random variable (i.i.d.) from the probability distribution obeyed by a single genotype at time $t$.
Thus, the variance of $\sum_{j} x_{i,j}$ is proportional to $N$, and then the variance of $y_i$ is proportional to $1/N$ because it is given by the summation of $x_{i,j}$ divided by $N$.
Therefore, the evolutionary rate should decrease if the number of chromosomes increases in the set inheritance mode.
Intuitively, it corresponds that an effect of a mutation in a single chromosome is divided by the number of chromosomes.
Indeed, in the numerical calculation, the variance of phenotype was proportional to $N^{-1}$ (Fig. \ref{fig:neutral_selection}B), and inverse of the first passage time, which is proportional to $D^{-3/2}$ \cite{Gardiner1985}, was proportional to $N^{-3/2}$ (Fig. \ref{fig:neutral_selection}C).
Additionally, even if we consider evolution on any fitness landscape instead of neutral evolution, the genetic variance of fitness should be smaller in polyploid organisms with the set inheritance mode because of the decrease in the phenotypic variance.
Hence, the evolutionary rate should be slower in any situation (Fig. \ref{fig:neutral_selection}E).

Contrarily, in the random inheritance mode, because a single chromosome may be duplicated multiple times, $x_{i,j}(t)$ may have a correlation for different values of $j$.
In this situation, the number of chromosomes is fixed, and each chromosome is neutrally selected.
This situation is similar to the gene fixation process through the neutral selection in a finite population, in which the single mutation is fixed within the same generation as the number of individuals in the population, on average \cite{Kimura1969}.
In the context of polyploidy, this indicates that all chromosomes are duplicated from one chromosome $N$ generations ago on average.
In other words, the number of generations in which all chromosomes are replaced by those replicated from a single chromosome equals the number of chromosomes possessed by the cell.
Thus, a set of chromosomes accumulates $N$ generations of mutations, and the variance of chromosomes is given as $N \sigma^2$ because an independent mutation with the variance $\sigma^2$ is applied to a chromosome in each generation (Fig. \ref{fig:neutral_selection}D).
Since randomly chosen chromosomes from the mother cell construct the chromosome set of the daughter cell, each $x_{i,j}$ is an i.i.d. from the probability distribution given by the mother cell's chromosome set.
Thus, the variance of $\sum_{j} x_{i,j}$ is proportional to $N^2 \sigma^2$, and then the variance of $y_i$ is proportional to $\sigma^2$ (Fig. \ref{fig:neutral_selection}B), which is independent of $N$.
Indeed, in the numerical calculation, the phenotypic variance and inverse of the first passage time were almost constant against changes in $N$ (Fig. \ref{fig:neutral_selection}C).
Note that the numerically calculated evolutionary rate was slightly slower, but this was because we set $x_{i,j}(0) = 0$ for all $i$ and $j$ and it takes $N$ generations for the variance of $x_{i,j}$ to relax.

Because the set and random inheritance modes represent the extreme cases, an organism with an imperfect DNA replication mechanism will exhibit an intermediate trait between two extreme modes.
This suggests that from the strong constraint of the inheritance manner, the genetic variance of the phenotype of polyploid organisms cannot exceed that of haploid organisms.
Indeed, the evolutionary rate of polyploid organisms on a smooth fitness landscape, which corresponds to the gradual evolution of quantitative traits, did not exceed that of the haploid organism with either the random or set inheritance mode (Fig. \ref{fig:neutral_selection}E).
Thus, polyploidy does not confer an advantage regarding evolution speed during neutral selection or gradual evolution.

\subsection{Evolution of novel traits can be accelerated in polyploid organisms}

\begin{figure}[htbp]
\centering
\includegraphics[width=0.8\linewidth]{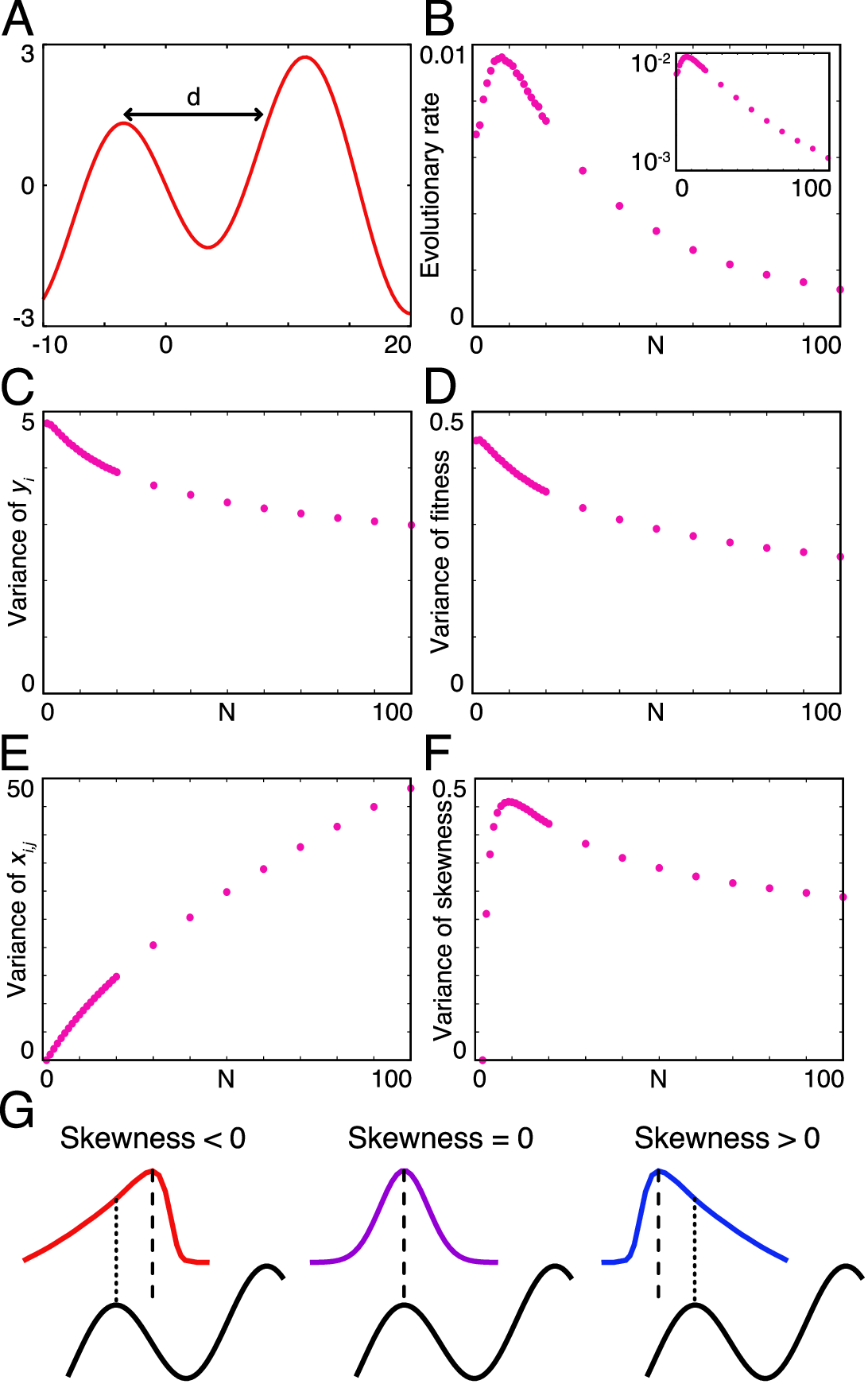}
\caption{
Evolution of a novel trait in polyploid organisms.
A) The multimodal fitness landscape given by $f(y) = \sin(y / 5) - 2 \sin(2 y / 5)$.
The local maximum is $y = -3.441212$, and the global minimum is $y = 11.374583$.
$d$ is the distance between a local maximal point and a distant point displaying the same fitness as the local maximum.
B) Evolutionary rate ($1/\langle \tau_t \rangle_{\mathrm{ens}}$) on the multimodal fitness landscape.
(Inset) Semi-log plot of the evolutionary rate.
C) Variance of phenotype in the population ($\langle (y_i - \langle y_i \rangle_{\mathrm{p}})^2 \rangle_{\mathrm{p}}$) at steady state.
We measured the behavior at steady state by setting the fitness of cells overflowed from a region between the local and global minimum as zero.
D) Variance of fitness in the population ($\langle (f(y_i) - \langle f(y_i) \rangle_{\mathrm{p}})^2 \rangle_{\mathrm{p}}$) at steady state.
E) Population average of variance of $x_{i,j}$ among chromosomes ($\langle \langle (x_{i,j} - \langle x_{i,j} \rangle_{\mathrm{c}})^2 \rangle_{\mathrm{c}} \rangle_{\mathrm{p}}$) at steady state.
F) Variance of skewness of $x_{i,j}$ (i.e., variance of $\langle (x_{i,j} - \langle x_{i,j} \rangle_{\mathrm{c}})^3 \rangle_{\mathrm{c}} / \langle (x_{i,j} - \langle x_{i,j} \rangle_{\mathrm{c}})^2 \rangle_{\mathrm{c}}^{3/2}$) at steady state.
G) Schematic representation of the effect of skewness on evolvability.
Colored lines reflect the distribution of $x_{i,j}$, and solid black lines represent fitness landscapes.
Dotted and dashed lines represent the mean and mode, respectively.
If skewness is zero, then the mode is consistent with the mean.
\label{fig:rugged_evo}}
\end{figure}

Next, to investigate whether polyploidy is not also advantageous outside of neutral selection and gradual evolution, i.e., for the evolution of novel traits, we introduced a simplified multimodal fitness landscape with global and local maximal peaks (Fig. \ref{fig:rugged_evo}A).
This is one of the simplest models for evolutionary innovation.
On the high dimensional fitness landscape, if one peak is connected to a second peak by the gradual ridgeline, on which the fitness slightly decreases and selection is almost neutral, then cells will evolve their traits through that line from one phenotype to another. 
Then, continuous quantitative traits will be observed between the two peaks, and no one will consider this situation as an appearance of evolutionary innovation.
Conversely, when deep valleys surround one peak in all directions, under which situation the fitness decreases lethally, cells must jump over a valley to reach another peak.
Then, the intermediate phenotype between two peaks is difficult to observe, and it corresponds well with the situation in which evolutionary innovation appears.
The multimodal fitness landscape presented in Fig. \ref{fig:rugged_evo}A is regarded as the projection of a path connecting two peaks of a high-dimensional fitness landscape onto one dimension.
The distance between the peaks is much larger than the mutational rate, which is the unique characteristic length in the phenotypic space determining the behavior of the model, and rare and drastic changes in genotype are required for a jump from one peak to another.

We assumed that all cells initially exhibited a locally but not globally optimal trait, i.e., $x_{i,j}(0)$ and thus, $y_i(0)$ was set as the value at the local maximum for all $i$ and $j$.
Then, we numerically calculated the time needed to jump to the maximal peak if we changed the number of chromosomes carried by the cells.
We first considered that all cells exhibited the random inheritance mode because the set inheritance mode of polyploidy decelerates the evolutionary rate regardless of the shape of the fitness landscape, as previously mentioned (Fig. \ref{fig:neutral_selection}).

When observing a single sample, the fitness of the fittest cell maintains the local maximum for some time and then suddenly jumps to the global maximum.
We defined the transition time $\tau_t$ when the average fitness of the population crosses a threshold (we set it as 1.5).
The jump from the local maximum to the global maximum follows a Poisson process, and the distribution of $\tau_t$ follows an exponential distribution.
Hence, the characteristic timescale for transition is described by the average of $\tau_t$, i.e., the evolutionary rate is represented by $1/\langle \tau_t \rangle_{\rm ens}$.

The evolutionary rate exhibited non-monotonic dependence on the number of chromosomes (Fig. \ref{fig:rugged_evo}B), as observed in the experiments.
When $N$ increased, the evolutionary rate increased, peaking around $N = 10$.
Then, the rate began to decline as an exponential of $-N$ (see the inset of Fig. \ref{fig:rugged_evo}B).
Thus, the existence of an optimal number of chromosomes for the evolutionary rate of novel traits was demonstrated in experiments and simulations.

To explain the existence of an optimal number of chromosomes for the evolutionary rate, we measured characteristics in steady state around the local maximum by setting the fitness of cells across the valley as zero.
If the number of chromosomes increases, the variance of $x_{i,j}$ almost linearly increases for small values of $N$, and the phenotypic variance and genetic variance of fitness monotonically decrease (Fig. \ref{fig:rugged_evo}C, D, and E).
Note that the variance of $x_{i,j}$ for large values of $N$ increases only sublinearly, in contrast to the case of neutral selection.
This may be because a cell with a genotype with excessively higher variance tends to fall into the valley and dies for large values of $N$.
This suggests that changes in the genetic variance of fitness are not sufficient to explain the dependence of the evolutionary rate on the number of chromosomes:
We cannot apply Fisher's fundamental theorem for natural selection to the evolution of novel traits in polyploidy organisms.

\subsection{Evolutionary innovation is described by the large deviation theory and it depends on the third-order moment of chromosomes}

The major difference between evolution under neutral selection and that of novel traits is the rarity of evolutionary events.
The former is obeyed by mundane mutations that accumulate over several generations, whereas the latter is obeyed by rare mutations introduced over a few generations.
Hence, we applied the large deviation theory to describe a rare event in which a daughter cell exhibits a largely different phenotype from the mother cell.

We considered the probability that a phenotype of a daughter cell crosses the valley between two phenotypes at the local and global maxima on the fitness landscape.
The chromosome set of a daughter cell was constructed as i.i.d. from the distribution of chromosomes in a mother cell, and a phenotype was given as the average of chromosomes.
In this analysis, we gave the distribution of the chromosome set of a mother cell with a mutation to the next generation as a cumulant-generating function $\log M(s; N)$, where $N$ is the number of chromosomes carried by the mother cell.
We denoted the distance in the phenotypic space between the local minimum and a distant point exhibiting the same fitness as $d$.
For the fixation of different phenotypes from the local maximum, a daughter cell must display higher fitness than that observed at the local minimum in the limit of high selection pressure.
Then, the daughter cell's phenotype must at least differ more than $d$ from the mother cell's phenotype. 
Such a probability of a rare event is known to follow the large deviation theory \cite{Varadhan1984} as follows:
\begin{equation}
P \left( \sum_{j=1}^{N'} \frac{x_{i,j}}{N'} \geq d \right) \simeq \exp \left( -N' I(d; N) \right),
\end{equation}
where $I(p; N)$ is a rate function as given by the Legendre transformation of a cumulant-generating function as $I(p; N) = \sup_{s} \left(s p - \log M(s; N) \right)$, where $N'$ is the number of chromosomes in the daughter cell.
Although the probability on the left-hand side converges to the value on the right-hand side when $N$ is infinite, the large deviation theory will give a valid approximation if $N$ is sufficiently large.
Therefore, if a cumulant-generating function is given, then the probability of the evolution of a novel trait can be calculated.

Because it is difficult to analytically derive the cumulant-generating function of a mother cell, we first assumed that it is given as the Gaussian distribution, which only has the first- and second-order cumulants.
The first-order cumulant is the average and the same as the phenotype of the mother cell.
Thus, the first-order cumulant is zero.
The second-order cumulant is the variance and a function of $N$, $f(N)$.
Then, the cumulant-generating function is given as
\begin{equation}
\log M(s; N) = \frac{f(N)}{2!} s^2,
\end{equation}
and the rate function is
\begin{align}
I(p; N) &= \sup_s \left( s p - \frac{f(N)}{2!} s^2 \right) \nonumber \\
&= \frac{p^2}{2 f(N)}.
\end{align}
Hence, the probability that the daughter cell achieves a novel trait is
\begin{equation}
P \left( \sum_{j=1}^{N'} \frac{x_{i,j}}{N'} \geq d \right) \simeq \exp \left( -N' \frac{d^2}{2 f(N)} \right).
\end{equation}
In this analysis, we considered the best situation in which the daughter cell can evolve a novel trait with the highest probability, where the variance $f(N)$ is proportional to $N$, as in the case of neutral selection, and it be given as $\sigma^2 N$.
Then, the aforementioned probability is given as
\begin{equation}
P \left( \sum_{j=1}^{N'} \frac{x_{i,j}}{N'} \geq d \right) \simeq \exp \left( -N' \frac{d^2}{2 N \sigma^2} \right).
\end{equation}
Because there is a constraint that mother and daughter cells carry the same number of chromosomes, i.e., $N'$ is equal to $N$, the aforementioned probability is calculated as
\begin{equation}
P \left( \sum_{j=1}^{N} \frac{x_{i,j}}{N} \geq d \right) \simeq \exp \left( -\frac{d^2}{2 \sigma^2} \right),
\end{equation}
and it is independent of $N$.
Thus, the evolutionary rate of novel traits is independent of the number of chromosomes.
Furthermore, $f(N)$ is sublinear of $N$ for large values of $N$ under selection. 
The aforementioned probability decreases with increasing $N$ as $\exp(-cN)$, consistent with the numerical result (the inset of Fig. \ref{fig:rugged_evo}B).
Therefore, if the distribution is Gaussian, that is, if we consider up to the second-order moment, then we never explain the acceleration of evolution in polyploid organisms.

We then considered the effect of the third-order moment.
We represented the skewness of distribution, the third-order cumulant normalized by 3/2 power of variance, as a function of $N$ as $g(N)$.
If the contribution of the third-order cumulant is smaller than that of the second-order one, then the probability that the daughter cell achieves a novel trait (see Supplementary Note 1 for derivation) is
\begin{equation}
P \left( \sum_{j=1}^{N'} \frac{x_{i,j}}{N'} \geq d \right) \simeq \exp \left\{ -N' \left( \frac{d^2}{2 f(N)} + \frac{8 d^3 g(N)}{3 f(N)^{\frac{3}{2}}} \right) \right\}. \label{eq:prob_skewness}.
\end{equation}
Then, when the sign of skewness differs from that of $d$, i.e., the mode of $x_{i,j}$ is closer to a peak corresponding to the novel trait than to the mean, then the probability of the rare event to across the valley increases (Fig. \ref{fig:rugged_evo}G).
Intuitively, it is interpreted that if a large bias of genetic information exists between chromosomes, then cells can hold a high frequency of mutations for novel traits without reducing their fitness.

To confirm the effect of skewness, we numerically calculated higher-order moment in steady state.
The population average of skewness was almost zero because the fitness landscape around the peak was near symmetry.
Then, we measured the variance of the skewness of chromosomes in the population to estimate the probability that a cell with high skewness emerges.
The result revealed non-monotonic dependency on $N$ similarly as observed for the evolutionary rate.
The variance of the skewness increased as $N$ increased for small values of $N$ and began to decline after the peak around $N \simeq 10$ (Fig. \ref{fig:rugged_evo}F).
From Eq. (\ref{eq:prob_skewness}), if the variance $f(N)$ is proportional to $N$, then the probability that the daughter cell achieve a novel trait is
\begin{equation}
P \left( \sum_{j=1}^{N} \frac{x_{i,j}}{N} \geq d \right) \simeq \exp \left( -\frac{d^2}{2 \sigma^2} - \frac{8 d^3 g(N)}{3 \sigma^{3} \sqrt{N}} \right).
\end{equation}
Thus, when the dependence of skewness on $N$ is a higher order than $\mathcal{O}(N^{\frac{1}{2}})$, the probability of achieving a novel trait increases with $N$, where $\mathcal{O}$ is the Landau symbol.
Although the variance of skewness is only a statistical amount of the whole population and not an amount for individual cells, an increase in variance indicates an increase in the number of cells with a large absolute value of skewness.
Hence, the agreement between the peaks of variance of skewness and the evolutionary rate implies that an increase in skewness in mother cells associated with an increase in the number of chromosomes accelerates the evolution of novel traits.

\begin{figure}[htbp]
\centering
\includegraphics[width=0.8\linewidth]{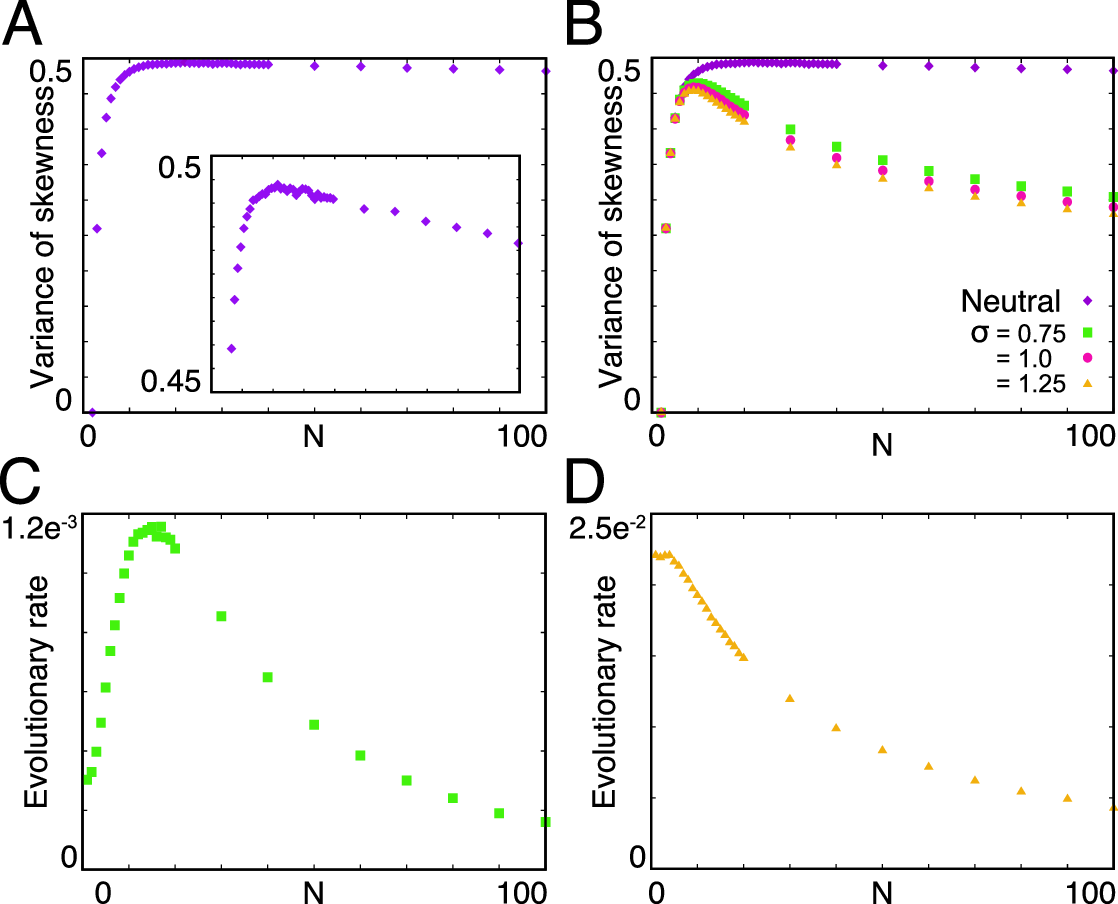}
\caption{
Dependence of the skewness of chromosomes on selection pressure.
A) Variance of skewness of $x_{i,j}$ at the time $t = 500$ under neutral selection.
(Inset) Enlarged plot of the variance of skewness.
B) Variance of skewness of cells with various mutation rates.
Purple diamonds: Neutral selection.
Green squares: $\sigma = 0.75$.
Red circles: $\sigma = 1.0$.
Yellow triangles: $\sigma = 1.25$.
C, D) Evolutionary rate of cells with mutation rates of (C) $\sigma = 0.75$ and (D) $\sigma = 1.25$.
\label{fig:mutation_rate}}
\end{figure}

Why does skewness exhibit non-monotonic dependency on the number of chromosomes?
Intuitively, in regions with small values of $N$, there are insufficient numbers of chromosomes for the distribution to be biased.
By contrast, in regions with large values of $N$, the distribution will be close to the Gaussian distribution because of the central limit theorem, and thus, the skewness must be small.
Indeed, even if the selection is neutral, the variance of the skewness displayed a maximal value of approximately $N \simeq 20 \text{--} 30$ (Fig. \ref{fig:mutation_rate}A).

If there is a selection process depending on the fitness, then a cell with a large $N$ and large skewness of $x_{i,j}$ will tend to fall into the valley and die.
Then, the skewness for large values of $N$ will be suppressed.
This implies that if the range of phenotypes, in which a cell is under nearly-neutral selection, is wider, then the optimal number of chromosomes exhibiting maximum skewness and evolutionary rates will become larger.
Indeed, when we set $\sigma$ as a smaller value, it corresponded to a wider range of nearly-neutral selection, both maximum skewness and the evolutionary rate became larger (Fig. \ref{fig:mutation_rate}B and C).
Although the value of $N$ at the peaks of the variance of skewness and the evolutionary rate slightly differed, this may be explained by the contribution of other moments.
For the case with a smaller mutation rate, a region in which the variance of $x_{i,j}$ linearly increases expanded for larger values of $N$ (see Supplementary Fig. \ref{fig:variance}), and thus, the peak of the evolutionary rate shifted to a larger $N$. 
Conversely, if the mutation rate is excessively large, then the evolution of novel traits is no longer a rare event, and thus, the large deviation theory cannot be applied.
Consequently, the haploid again has an advantage regarding the speed of evolution, as it exhibits the highest variance in phenotype (Fig. \ref{fig:mutation_rate}B and D).
Note that the normalized fourth-order moment kurtosis did not exhibit non-monotonic dependence on the number of chromosomes, and it was almost independent of the mutation rate (Supplementary Fig. \ref{fig:kurtosis}).
Hence, only skewness displayed non-monotonic dependence on the number of chromosomes at least up to fourth-order moment.

\section{Discussion}

In this study, we demonstrated that polyploidy confers the advantage of evolving novel traits even though the genetic variance of the phenotype does not increase with polyploidy.
To elucidate this seemingly contradictory problem to Fisher's fundamental theorem of natural selection, we theoretically analyzed rare evolutionary events by introducing the large deviation theory.
We found that the skewness of the distribution of genotypes coded in each chromosome is essential to determining the probability of evolving new traits.
Because skewness peaked with a finite number of chromosomes (Fig. \ref{fig:rugged_evo}F), there is an optimal number of chromosomes for evolving novel traits (Fig. \ref{fig:rugged_evo}B).

\begin{table*}[]
    \centering
    \caption{Typical number of chromosomes in each cyanobacterial species}
    \begin{tabular}{l | c c}
    Species & Number of chromosomes & Ref. \\
    \midrule
    \textit{Synechococcus elongatus} PCC 7942 & 2--6 & \cite{Chen2012, Jain2012, Zheng2017, Ohbayashi2019} \\
    \textit{Synechococcus} sp. PCC 6301 & 2--8 & \cite{Binder1990} \\
    \textit{Anabaena variabilis} & 8--9 & \cite{Simon1980} \\
    \textit{Anabaena} sp. PCC 7120 & 8.2 & \cite{Hu2007} \\
    \textit{Microcystis} sp. & 1--10 & \cite{Kurmayer2003} \\
    \textit{Synechococcus} sp. PCC 7002 & 5--11 & \cite{Moore2019, Ohbayashi2020} \\
    \textit{Synechocystis} sp. PCC 6803 & 9.7--22.2 & \cite{Zerulla2016} \\
    \textit{Anabaena cylindrica} & 25 & \cite{Simon1977} \\
    \textit{Cyanobacterium aponinum} PCC 10605 & 16--32 & \cite{Ohbayashi2020} \\
    \textit{Geminocystis} sp. NIES-3708 & 17--34 & \cite{Ohbayashi2020} \\
    \bottomrule
    \end{tabular}
    \label{tab:chromosomes}
\end{table*}

Although the optimal number of chromosomes depends on the shape of the fitness landscape and the mutation rate, the upper limit of the optimal number can be estimated as approximately 20 -- 30 because the maximal skewness is approximately $N = 20 \text{--} 30$ in the case of neutral selection, and it decreases under selection.
Interestingly, such a range of the number of chromosomes agrees well with the actual number of chromosomes in various species of cyanobacteria (see Table \ref{tab:chromosomes}).
This suggests that the number of chromosomes in cyanobacteria might result from optimization for evolutionary innovation.
Moreover, we found that the optimal number of chromosomes is correlated with the range of the fitness landscape in which a cell can change its phenotype under nearly-neutral selection.
Therefore, further cooperative studies involving theoretical, experimental, and bioinformatic analyses will help us infer the fitness landscape of polyploid organisms based on the number of chromosomes.

Because our model and settings are general, our theory will be universally applicable to other polyploid organisms than cyanobacteria.
In particular, our theory will be useful for organisms without sophisticated DNA replication control mechanisms.
PGCCs and L-form bacteria represent promising examples of the successful application of this theory.
Polyploidy in PGCCs is attributable to errors in the chromosome replication and segregation mechanism, and thus, random chromosome replication and segregation are expected to occur opposed to normal replication and segregation.
Similarly, random chromosome inheritance is expected to occur in L-form bacteria, which develop polyploidy through cell fusion.
Because our theory is applicable regardless of the details of the systems as long as chromosomes are inherited in random mode, we believe that our theory can unravel the mechanisms of achieving novel drug resistance and metastatic potential in cancer cells and multidrug-resistant bacteria.

In addition, we expect that our results can be applied to cases of gene conversion \cite{Bittihn2017} and multi-copy plasmids \cite{Rodriguez-Beltran2018}, both of which were previously reported to be capable of facilitating the evolutionary innovation.
In particular, the typical number of native plasmids ranges from approximately 10 to several dozen \cite{Yano2019}, which corresponds to the maximal skewness observed in the present study (Fig. \ref{fig:mutation_rate}A).
This suggests that the copy number of native plasmids might result from optimization for evolutionary innovation.
We expect that applying our theory to plasmid systems and experiments in such systems will uncover the evolutionary advantage of the regulation of plasmid copy numbers.

Furthermore, our study has provided a new theoretical perspective on neofunctionalization by whole-genome duplication as described by Susumu Ohno \cite{Ohno1970}.
Considering the cells immediately after whole-genome duplication, they would not have had sophisticated duplication mechanisms.
Thus, our theory may be applicable to whole-genome duplication events, and the skewness of chromosomes may provide the theoretical representation of the early phase of neofunctionalization.
This study paves the way for the understanding of evolution via whole-genome duplication, and the mechanism by which that early biased segregation of genes, as represented by skewness, is fixed is an important future problem to understand the late phase of neofunctionalization.
We expect that additional theoretical studies with more complex models, bioinformatic models, and further constructive experimental studies would perfectly reveal the mechanism of neofunctionalization in the future.

\begin{acknowledgments}
We thank Kunihiko Kaneko and Chikara Furusawa for a critical reading of the manuscript, and Tomoko Ohta for fruitful discussion.
\end{acknowledgments}

\appendix

\section{Derivation of the rate function with third-order cumulant.}

A cumulant generating function up to the third-order term is given as
\begin{equation}
\log M(s) = \mu s + \frac{\sigma^2}{2!} s^2 + \frac{c_3}{3!} s^3 \label{eq:cumulant},
\end{equation}
where $\mu$, $\sigma^2$, and $c_3$ are the average, variance, and the third-order cumulant, respectively. 
Here, we set $\mu$ as zero for simplicity without loss of generality. The rate function for the above cumulant generating function is
\begin{align}
I(p) &= \sup_s \left( sp - \log M(s) \right) \nonumber \\
&= \sup_s \left( s p - \frac{\sigma^2}{2!} s^2 - \frac{c_3}{3!} s^3 \right).
\end{align}
Since the first derivative of $sp - \log M(s)$ is
\begin{equation}
\frac{d \left( sp - \log M(s) \right)}{ds} = p - \sigma^2 s + \frac{c_3}{2} s^2,
\end{equation}
and $s$ for the extrema is
\begin{equation}
s = \frac{-\sigma^2 \pm \sqrt{\sigma^4 + 2 p c_3}}{c_3}.
\end{equation}
The second derivative of $sp - \log M(s)$ is
\begin{align}
\frac{d^2 \left( sp - \log M(s) \right)}{ds^2} &= -\sigma^2 - c_3 s \nonumber \\
&= -\sigma^2 + \sigma^2 \mp \sqrt{\sigma^4 + 2 p c_3} \nonumber \\
&= \mp \sqrt{\sigma^4 + 2 p c_3},
\end{align}
and then,
\begin{equation}
s^* = \frac{-\sigma^2 + \sqrt{\sigma^4 + 2 p c_3}}{c_3}
\end{equation}
is $s$ for the maximal value when $\sigma^4 + 2 p c_3$ is $\geq 0$.
Hence, the rate function is given as
\begin{widetext}
\begin{align}
I(p) &= s^* p - \frac{\sigma^2}{2!} s^{*2} - \frac{c_3}{3!} s^{*3} \nonumber \\
&= \frac{2 p^2}{\sigma^2 + \sqrt{\sigma^4 + 2 p c_3}} - \frac{2 p^2 \sigma^2}{\left( \sigma^2 + \sqrt{\sigma^4 + 2 p c_3} \right)^2} - \frac{4 c_3 p^3}{3 \left( \sigma^2 + \sqrt{\sigma^4 + 2 p c_3} \right)^3} \nonumber \\
&= \frac{2 p^2}{\sigma^2} \frac{1}{\left( 1 + \sqrt{1 + \frac{2 p c_3}{\sigma^4}} \right)^2} + \frac{8}{3} \frac{c_3 p^3}{\sigma^6} \frac{1}{\left( 1 + \sqrt{1 + \frac{2 p c_3}{\sigma^4}} \right)^3}. \label{eq:rate_function_full}
\end{align}
\end{widetext}
In the limit of $c_3 \rightarrow 0$, the above equation is
\begin{equation}
I(p) = \frac{p^2}{2 \sigma^2},
\end{equation}
which coincides with Eq. (3) in the main text, a case of Gaussian distribution.
When the squared variance is much larger than the third-order cumulant, $|c_3| \ll \sigma^4$, Eq. (\ref{eq:rate_function_full}) is
\begin{align}
I(p) &\simeq \frac{2 p^2}{\sigma^2} + \frac{8}{3} \frac{c_3 p^3}{\sigma^6} \nonumber \\
&= \frac{2 p^2}{\sigma^2} + \frac{8}{3} \beta \frac{p^3}{\sigma^3}, \label{eq:rate_function}
\end{align}
where $\beta$ is skewness, given as $\beta = \frac{c_3}{\sigma^3}$.
When the variance and skewness depend on the number of chromosomes $N$, and are given as functions of $N$, $\sigma^2 = f(N)$ and $\beta = g(N)$, respectively, the rate equation is
\begin{equation}
I(p; N) = \frac{2 p^2}{f(N)} + \frac{8}{3} g(N) \frac{p^3}{f(N)^{3/2}},
\end{equation}
and the probability that the daughter cell achieves a novel trait is
\begin{equation}
P \left( \sum_{j=1}^{N'} \frac{x_{i,j}}{N'} \geq d \right) \simeq \exp \left\{ -N' \left( \frac{d^2}{2 f(N)} + \frac{8 d^3 g(N)}{3 f(N)^{\frac{3}{2}}} \right) \right\}.
\end{equation}

\renewcommand{\figurename}{Supplementary FIG.}
\setcounter{figure}{0}

\begin{figure}[htbp]
\centering
\includegraphics[width=0.5\linewidth]{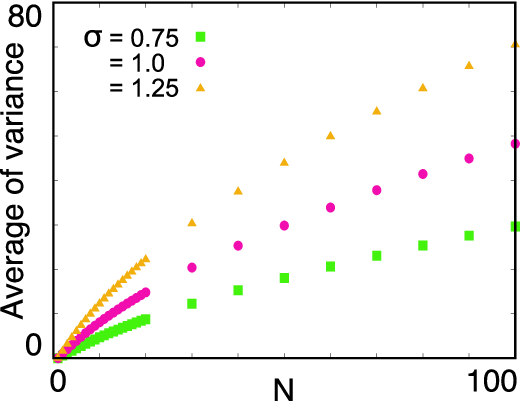}
\caption{
Dependence of the variance of chromosomes on the mutation rate.
Average of variance of each cell over the population is plotted.
Green squares: $\sigma = 0.75$.
Red circles: $\sigma = 1.0$.
Yellow triangles: $\sigma = 1.25$.
}
\label{fig:variance}
\end{figure}

\begin{figure}[htbp]
\centering
\includegraphics[width=0.5\linewidth]{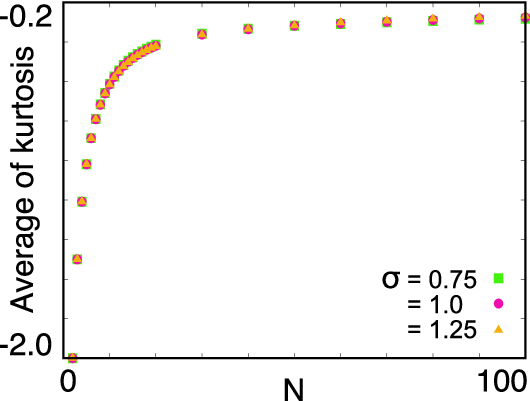}
\caption{
Dependence of the kurtosis of chromosomes on the mutation rate.
The average of kurtosis of each cell over the population is plotted.
Green squares: $\sigma = 0.75$.
Red circles: $\sigma = 1.0$.
Yellow triangles: $\sigma = 1.25$.
}
\label{fig:kurtosis}
\end{figure}

\end{document}